**Revealing Information from Weak Signal in Electron Energy-Loss Spectroscopy with a Deep Denoiser**

Yifan Wang[1], Mai Tan[1], Carlos Fernandez-Granda[2], Peter A. Crozier[1]

1. Materials Science and Engineering, School for Engineering of Matter, Transport and Energy, Arizona State University, Tempe, AZ, USA

2. Courant Institute of Mathematical Sciences, New York University, New York, NY, USA

**Abstract**

Electron energy-loss spectroscopy (EELS) coupled with scanning transmission electron microscopy (STEM) is a powerful technique to determine materials composition and bonding with high spatial resolution. Noise is often a limitation especially with the increasing sophistication of EELS experiments. The signal characteristics from direct electron detectors provide a new opportunity to design superior denoisers. We have developed a CNN based denoiser, the unsupervised deep video denoiser (UDVD), which can be applied to EELS datasets acquired with direct electron detectors. We described UDVD and explained how to adapt the denoiser to energy-loss spectral series. To benchmark the performance of the denoiser on EELS datasets, we generated and denoised a set of simulated spectra. We demonstrate the charge spreading effect associated with pixel interfaces on direct electron detectors, which leads to artifacts after denoising. To suppress such artifacts, we propose some adjustments. We demonstrate the effectiveness of the denoiser using two challenging real data examples: mapping Gd dopants in $CeO_2$ nanoparticles and characterizing vibrational modes in hexagonal boron nitride (h-BN) with atomic resolution.

# 1. Introduction

Electron energy-loss spectroscopy (EELS) is a powerful technique for materials characterization, providing information on elemental, electronic, plasmonic, photonic, and phononic properties of materials (Bosman et al., 2007; Egerton, 2011; García de Abajo and Kociak, 2008; Krivanek et al., 2014; Liu et al., 2019; Nelayah et al., 2007; Venkatraman et al., 2019; Wang et al., 2023). When combined with scanning transmission electron microscopy (STEM), EELS can achieve atomic scale spatial resolution (Pennycook and Nellist, 2011). The energy-loss spectrum spans a wide range of intensity levels, with signal strength varying depending on the sample and the type of information sought. For instance, the valence loss region often exhibits strong plasmon excitations, whereas inner-shell and vibrational spectroscopy signals are typically much weaker.

As EELS experiments become increasingly sophisticated and ambitious, there is a continuous push to map larger areas, achieve higher sensitivity, and reduce acquisition times. While detector technology continues to advance (Levin, 2021), the precision of EELS measurements is often limited, not by detector hardware, but by the inherent signal-to-noise ratio (SNR) of the electron signal of interest. In raw spectral data, the highest SNR is fundamentally determined by shot noise, which arises from the Poisson statistics governing electron emission and scattering processes. The SNR can be improved by increasing the electron probe current but in many cases this is often constrained by the radiation sensitivity of the sample (Egerton, 2019). The SNR can also be improved by extending the acquisition time; however, in practice the acquisition times are limited by instrument stability.

The ability to detect a feature in a single noisy spectrum (or image) is traditionally assessed using the Rose criterion, which states that the signal from the feature of interest must exceed the noise in the total signal by a significant margin (Rose, 1948). However, when prior system knowledge or large volumes of data are available, statistical methods can be applied to surpass this threshold, leading to the development of advanced noise reduction techniques. Over the years, various approaches to noise reduction in spectral signals have been explored, ranging from simple averaging methods to more sophisticated techniques such as principal component analysis (PCA) (Bosman et al., 2007). Recently, neural network (NN) methods have emerged as powerful tools for denoising scientific data (Pate et al., 2021).

Previously, we developed a denoiser based on convolutional neural networks (CNN), namely the unsupervised deep video denoiser (UDVD) (Sheth et al., 2021; Marcos-Morales et al., 2023). UDVD is based on a deep convolutional neural network, which is trained exclusively on noisy data, exploiting the low spatial correlation of the noise in direct electron detectors (Levin, 2021). In our previous report, UDVD was successfully applied to noisy *in situ* TEM movies, revealing the evolution of the surface atomic structure of Pt nanoparticles in a gas environment at a high spatial-temporal resolution (Crozier et al., 2025). Since direct electron detectors are also used as EELS cameras, denoising EELS datasets is a natural application of UDVD.

In this manuscript, we propose and explore the application of the UDVD to spectral datasets acquired on a direct electron detector. We briefly describe UDVD and explain how to adapt the denoiser to energy-loss spectral series. To benchmark the performance of the denoiser on EELS datasets, we generated and denoised a set of simulated spectrum images. We present the effect of low SNR on the final denoising result. We demonstrate the charge spreading effect associated with pixel interfaces on direct electron detectors, which leads to artifacts after denoising. To suppress such artifacts, we propose some adjustments. We demonstrate the effectiveness of the denoiser using two challenging real data examples: mapping Gd dopants in $CeO_2$ nanoparticles and characterizing high momentum transfer phonon modes in hexagonal boron nitride (h-BN).

## 2. Methods

*2.1 Unsupervised deep video denoiser (UDVD) and adaptation to EELS*

UDVD is based on the deep-learning framework for denoising, where a deep convolutional neural network is trained to remove noise automatically by optimizing its performance on a training dataset. In order to enable training only with noisy data, UDVD leverages the "blindspot" mechanism (Batson, 1982; Krull et al., 2019; Laine et al., 2019). The network produces an output that approximates its noisy input. In order to avoid overfitting the noise and learning the trivial identity function, which maps the noisy input directly to the output, the neural-network architecture is designed so that the network estimates each noisy pixel *without having access to that same noisy pixel*. As a result, it must estimate each pixel using only the remaining data. If the noise is pixel-wise uncorrelated, this precludes overfitting, as the network cannot approximate the unobserved noise.

The UDVD architecture takes 5 frames of 2D data as input (**fig 1a**). The five frames are rotated 0°, 90°, 180° and 270° (R) in four branches. Each branch consists of two Unet-style (Ronneberger et al., 2015) components (D1 and D2), with shared parameters across branches. These components contain vertically causal convolutional filters, i.e., the convolution operation only considers information from above the current pixel position. The output from each branch is derotated ($R^{-1}$) and combined linearly (1x1 convolution) to form the denoised output. Due to the vertically causal filters, the rotation and the derotation process implements the desired blindspot mechanism, which excludes the pixel of interest from the receptive field of the networks (**fig 1b).** During training, the parameters in UDVD are optimized by minimizing the mean squared error (MSE) between its output and the noisy data via stochastic gradient descent. The training process is consequently unsupervised and only requires the input noisy datasets.

It should be noted that an important assumption underlying UDVD (and any blindspot-based unsupervised denoising method) is that the noise is spatially uncorrelated. Otherwise the denoiser will learn to estimate the noise from the input, instead of removing it. Because of this, fixed pattern noise and flicker noise from an electron source may be perceived as signal by the denoiser. More details about the denoiser can be found in our previous work (Sheth et al., 2021; Marcos-Morales et al., 2023; Crozier et al., 2025; Mohan et al., n.d.). The code of the denoiser can be found on GitHub (https://github.com/crozier-del/UDVD-MF-Denoising ).

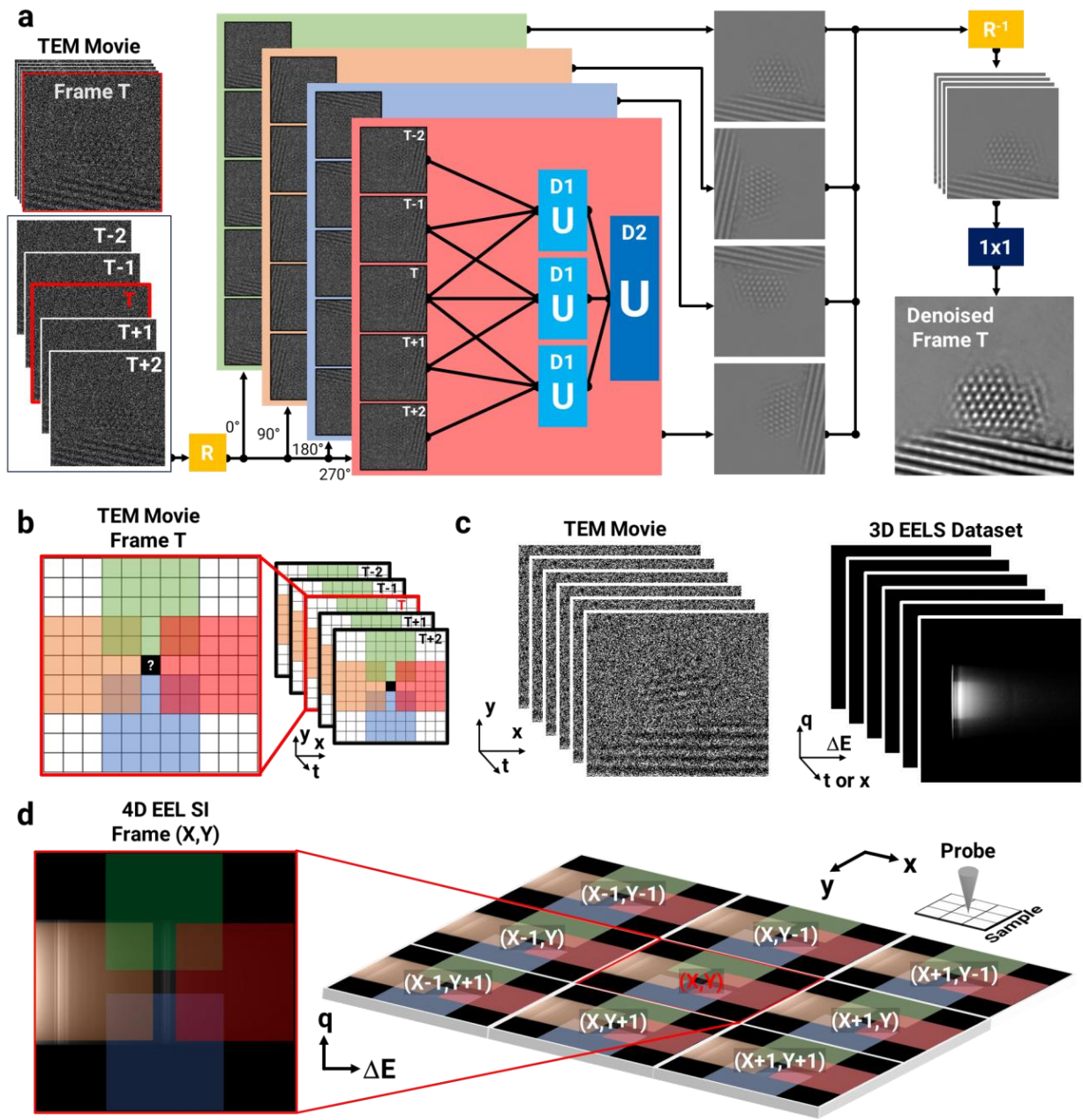

**Figure 1:** UDVD architecture with EELS data input. **a.** Schematic diagram of UDVD. **b.** The receptive field (colored regions) for pixel of interest (black pixel with white question mark) used by UDVD. The colored box represents the receptive field of the corresponding component in **a**. **c.** Comparison between TEM movies and 3D EELS datasets. For a series of EELS datasets, the EELS frames are perceived by the denoiser as the frame dimension. **d.** Adaptation of the blind spot architecture to 4D EEL SI dataset.

| Data Type | Data Dimension | Frame Dimension | Sequence Dimension |
|---|---|---|---|
| TEM Movie | $3D, (x, y, t)$ | $(x, y)$ | $t$ |
| Time-Series EELS | $3D, (\Delta E, q, t)$ | $(\Delta E, q)$ | $t$ |
| EELS Line-Scan | $3D, (\Delta E, q, x)$ | $(\Delta E, q)$ | $x$ |
| EEL Spectrum Image | $4D, (\Delta E, q, x, y)$ | $(\Delta E, q)$ | $(x, y)$ |

**Table 1:** Dimensions of different types of TEM datasets.

UDVD was originally designed to denoise movies and was demonstrated to work on natural photographic images, bright-field phase contrast TEM images and other microscopy modalities (Sheth et al., 2021). In order to apply it to EELS data, we reorganize the data in such a way that it can be input into the existing UDVD architecture with minimal changes. As **table 1** shows, the dimension of a TEM movie dataset can be separated into two parts: the frame dimension (TEM images $(x, y)$), and the sequence dimension (time $(t)$). Compared to TEM movies, EELS datasets have a different structure (**fig 1c**). The raw data of a time-series EELS captured on the spectrometer detector has three dimensions: energy loss $(\Delta E)$, non-disperse/momentum transfer $(q)$, and time $(t)$. We reorganize the data so that the two first dimensions form a 2D spectrum frame $(\Delta E, q)$ and the sequence dimension is the time (t). For notational convenience, the spectrum in 2D format will be referred to as **EELS frame** for the rest of this manuscript. EELS line-scan datasets have a similar structure to time-series EELS, but the time axis $t$ is replaced with the spatial axis $x$. Electron energy-loss spectrum images (EEL SI) have two spatial axes $x$ and $y$. The sequence coordinate is associated with x and y positions of the electron probe, thus resulting in a 4D dataset (see **table 1**). In order to apply UDVD, we reorganize the EEL SI data by considering both spatial directions as the sequence direction (**fig 1d**). In addition, we incorporate additional input frames (9 in total), as explained in detail in **supplement S1**.

*2.2 Microscopy and Denoising*

All the datasets showcased here were acquired on a Nion UltraSTEM 100 equipped with a monochromator and a Dectris ELA hybrid-pixel detector, with dimensions of 1028 × 512. Spatially resolved spectrum image (SI) datasets were collected as EELS frames, resulting in 4D datasets. The denoised 4D datasets were integrated along the q axis to retrieve the 1D spectrum, thus forming a 3D SI. The denoiser script was run on ASU Sol HPC, using a Nvidia A100 80Gb. For a dataset of 32 × 32 EELS frames, it takes 2 hours and 15 minutes to train UDVD for 50 epochs, and apply it to denoise the data.

*2.3 Evaluation of Denoising Results on Simulated Dataset*

In order to evaluate the denoiser, we divided the simulated data at random into a training set (70%) and a test set (30%). Once the denoiser is trained on the training set, we evaluate it on the test set using the peak signal-to-noise ratio (PSNR) (Nadipally, 2019, p. 2):

$$PSNR = 10 \cdot \log_{10}\left(\frac{MAX_I^2}{MSE}\right) \qquad (2)$$

where $MAX_I, MSE$ are the maximum value of the regional of the interest, and the mean squared error between the denoised estimate and the corresponding ground truth data, respectively. Since the spectrum does not fully cover the whole detector, the top and bottom regions of the EELS frame are zero. Thus, those regions are excluded from the calculation of the PSNR to reflect the effective value of PSNR. It should be noted that the PSNR value is normalized using the maximum value of ground truth data, so that PSNR values of datasets that have different signal strengths can be compared to each other. An increase in PSNR of $x$ dB corresponds to increasing the signal by $10^{0.1x}$ times. (See **supplement S2**)

## 3. Results and Discussion

*3.1 Denoising Simulated Core-loss Dataset*

To benchmark the performance of the denoiser, simulated noisy EELS maps were generated from BN structures containing B and N point defects in the form of vacancies and adatoms (See **supplement S3** for the method of generation), and then denoised. **Fig 2** shows the results from a dataset with a beam current of 10 pA, and a pixel time of 4 ms (corresponding to $2.5*10^5$ electrons). **Fig 2a** shows a comparison between the raw, denoised, and ground truth EELS frames. Line features, i.e., elemental edges, can be seen in the denoised EELS frames, corresponding to features in the ground truth image, while they are hardly visible in the noisy raw data. To retrieve the 1D spectrum, the EELS frames in **fig 2a** are integrated along the non-dispersive axis, and the result of the integration is shown in **fig 2b**. Compared to the ground truth, the features in the denoised result have similar shape and intensity, but the peaks are broadened by ~3 eV for B K-edge, and ~6 eV for N K-edge. The broadening of the signal can be considered as losing details of the clean signal. With higher SNR, the neural network can preserve finer details in the. In this case, the SNR is not strong enough to determine some of the details, therefore causing the peak broadening.

The SIs were generated using the same energy window to fit the background. As **fig 2c** shows, the raw SI merely resembles the atomic structure, while the subtle difference caused by extra adatoms or vacancies can be seen in the denoised SI, marked with arrows. The raw dataset has noisier background, which makes background subtraction more challenging, i.e., random fluctuations in the dataset can lead to different background fittings. However, since the denoised dataset has smoother background (**fig 2b**), the background subtraction is more stable across the whole map. Thus, subtle differences can be resolved with higher certainty in the denoised dataset. The results also demonstrate the capability of the denoiser to detect minor changes in the spectrum.

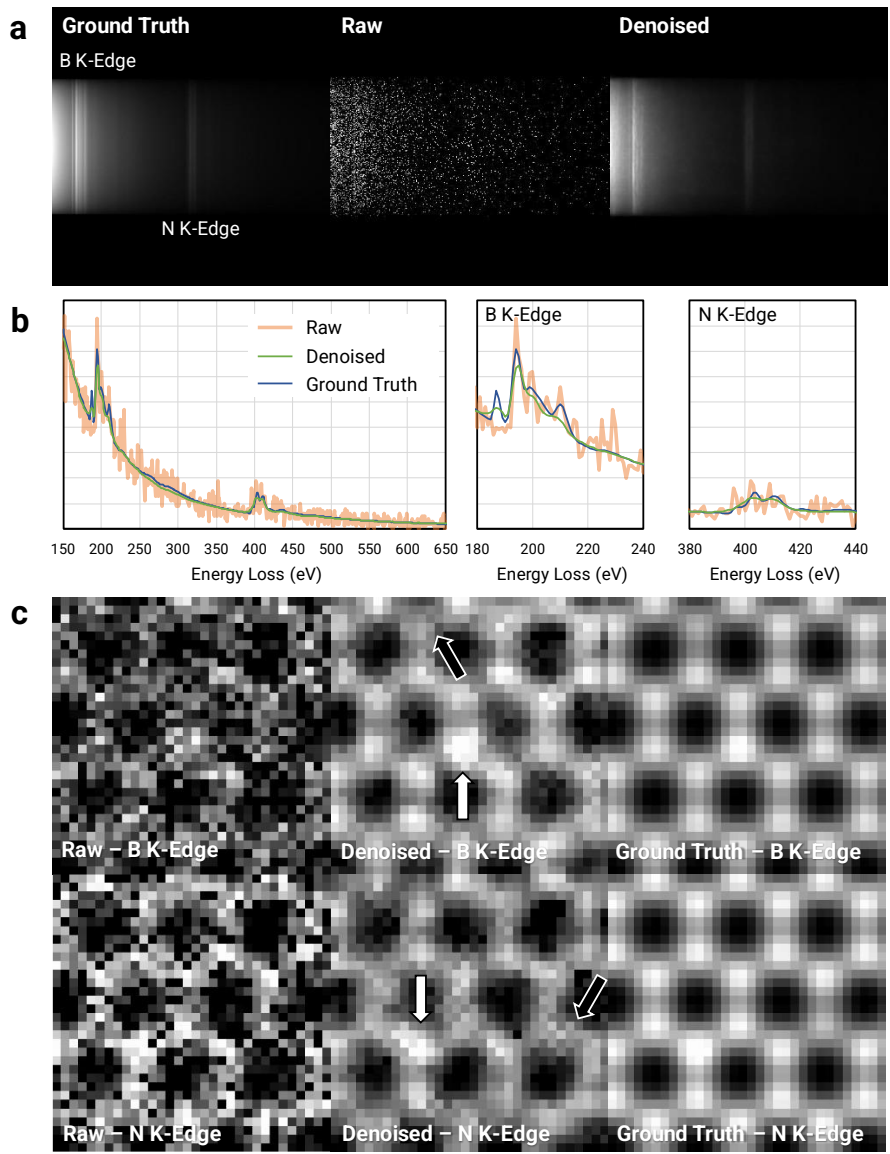

**Figure 2:** Denoising result of simulated h-BN core-loss dataset. **a.** Comparison among the ground truth, raw, and denoised EELS frames. All images correspond to the same point. The brightness and contrast are fixed across the different images. The elemental edges are marked in the ground truth EELS frame. **b.** Comparison among raw (orange), denoised (green), and ground truth (dark blue) 1D spectrum. The images in **fig 2a** are summed along the non-dispersive axis, resulting in a 1D spectrum. The B K-edge and N K-edge features are magnified and demonstrated on the right side. **c.** Comparison among raw, denoised, and ground truth background subtracted SIs. Images in the top row are B K-edge (190 eV – 210 eV) maps, and images in the bottom row are N K-edge (400 eV – 420 eV) maps. The extra atoms and vacancies are marked with white and black arrows in the denoised SI, respectively.

To explore the performance of the denoiser on simulated raw data of various signal levels, the beam current was kept constant at 10 pA, and the pixel time was set from 1ms to 64 ms. **Fig 3a-b** demonstrates the PSNR change with signal strength (since the signal strength is linearly proportional to pixel time). The PSNR of denoised results are always higher than the corresponding raw dataset. For the EELS frame, the denoiser improves the PSNR by more than 30 dB, while for the 1D spectrum, the denoiser improves the PSNR by more than 10 dB, i.e., the signal strength of the raw data has to be 10 times higher to match the denoised result. However, as both **fig 3a** and **b** show, the increment of PSNR is not constant. For the raw data, it increases logarithmically (the x-axis is in logarithmic scale). When the signal strength increases by 2, the $MAX_I$ and $MSE$ of the signal also increase by 2. Therefore, the increase in PSNR is ~ 3.01 as calculated from **eq 2**, which matches the graphs. In addition to calculating the full spectrum PSNR, it is helpful to calculate PSNR for the B K and N -K-edges. The change in the PSNR with dose for the full spectrum and B K edge is similar, while the trend of N-K edge PSNR is slightly different, especially in the 1D spectrum **(fig 3b)**. The main reason for this is that the B K-edge has a much stronger signal (including the background) than the N K-edge. Thus, B K-edge will contribute more to the MSE. In other words, the full spectrum PSNR is dominated by the B K-edge.

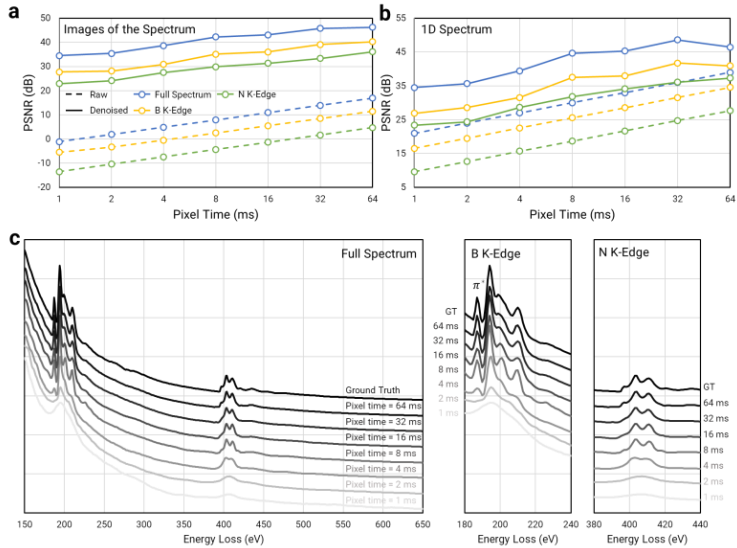

**Figure 3** Denoising result of raw data with different signal strength. **a-b.** PSNR as a function of pixel time. PSNR of the full spectrum, B K-edge (180 eV – 240 eV), and N K-edge (380 eV – 440 eV) are marked with blue, yellow, and green circles and lines, respectively. Dashed lines correspond to raw dataset, and solid lines correspond to denoised dataset. The x axis is in logarithmic scale. **a.** PSNR of EELS frames. **b.** PSNR of 1D spectrum. **c.** Comparison among denoised results with different pixel time and the ground truth (GT). All the spectra are normalized and shifted in y axis. The B K-edge and N K-edge features are magnified and demonstrated on the right side.

To link the PSNR to the actual spectrum, the normalized spectrum of denoised results and the ground truth are shown in **fig 3c**. By comparing the denoised spectra to the ground truth spectra (right side of **fig 3c**), we find that for B K-edge, dataset with pixel time more than 8 ms resembles the ground truth, i.e., all the major peaks in the near-edge fine structure appear with near correct intensity ratios. While for N K-edge, the spectra resemble ground truth for pixel time of 32 ms and 64 ms. The PSNR of elemental edges in the 1D spectra are all above 35 dB. Therefore, for a region that mostly consists of strong features, a PSNR value of 35 can be considered as a threshold for retrieving the fine structure of the edge. Also, as the figure shows, when the signal is weak, e.g., 1 ms or 2 ms pixel time for both B and N K-edge, the fine structures of the edge are perceived by the denoiser as one broad peak. As the signal becomes stronger, more details can be retrieved by the denoiser, e.g., the π* peak of boron K-edge. As we described above, the capability to differentiate fine details of the clean signals and noise of the neural network depends on the raw data SNR. In other words, the denoiser needs more signal to retrieve complicated near-edge fine structure features.

*3.2 Detector Considerations: The Effect of Charging Spreading and Pixel Interfaces on Denoising*

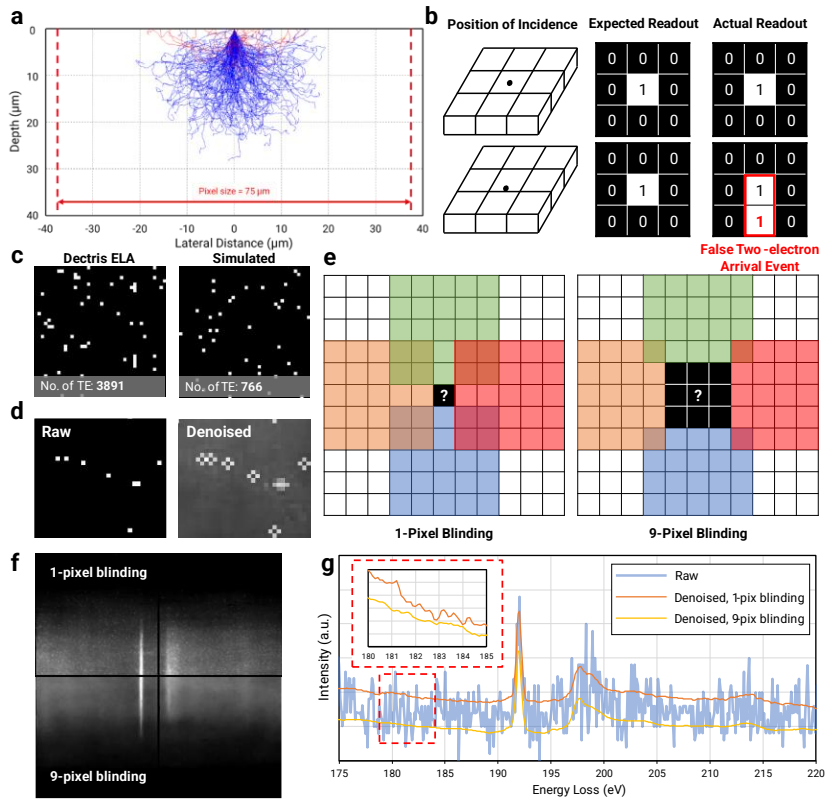

**Figure 4:** Signal spreading in direct electron detector **a.** The Monte Carlo simulation of 2000 electrons incident in Si. The blue lines represent the trajectories of forward scattering electrons, and the red lines are trajectories of back-scattered electrons. The red dashed line marks the lateral size of a pixel on Dectris ELA

detector. **b.** Schematic diagram of the detector's response. The red box shows an example of the non-zero pair. **c.** Comparison of a simulated dataset with an experimental dataset. The bright pixels represent non-zero values (mostly 1), and dark pixels represent 0. **d.** Effect of signal spreading on the neural network. The left and right image are the same area from the raw and the denoised dataset. For the raw image, bright pixels represent 1 and dark pixels represent 0. **e.** Schematic diagram of modification to blinding. The left one is the original blinding algorithm (**1-pixel bslinding**), and the right one includes neighboring pixels in the blinded region (**9-pixel blinding**). The black boxes represent blinded pixels that will not be included in the receptive field of the neural network. The colored boxes are the receptive field of the neural network used to estimate the value of the pixel marked by question mark. **f.** Comparison between denoised data using the 1-pixel blinding scheme (top half), and denoised data using the 9-pixel blinding scheme (bottom half). The vertical and horizontal dark lines are the chip boundaries on the detector. The brightness and contrast are set the same. The cross-shape artifacts are visible as bright speckles. **g.** Comparison between raw data (blue line), denoised data using the 1-pixel blinding scheme (orange line), and denoised data using the 9-pixel scheme (yellow line). Data was shifted vertically for the convenience of comparison. The figure inset is a magnified version of the comparison of orange and yellow line without the shift, marked by the red dashed box.

When a fast electron impacts a direct electron detector, the energy is deposited in the form of electron-hole pairs (along the trajectory), which are collected and converted into the signal (Clough and Kirkland, 2016). **Fig 4a** demonstrates 2000 trajectories of 60 keV electrons in silicon (a common material for sensors) simulated with Monte Carlo method (Drouin et al., 2007). As the trajectory shows, when an electron hits a silicon detector, it can travel more than 20 μm in lateral dimension. In other words, the generated signal can be spread widely in the lateral dimension. If the electron arrives close to the interface between two pixels, there will be a significant probability of generating signal in both pixels (**fig 4b**), giving a false indication that two electrons have arrived, namely false adjacent two-electron arrival events (false TE).

The spread of the signal in the detector silicon in a sparse dataset can be demonstrated through the number of TEs. Such spread will increase the number of the TEs compared to a detector with no correlation between pixels, since the latter can only have TEs by having two electrons arriving at adjacent pixels. Assuming the process follows a Poisson distribution, i.e., for a given mean dose of $\lambda$ electrons per pixel, the probability ($P$) of $k$ electrons arriving at one pixel is:

$$P(x = k) = \frac{e^{-\lambda}\lambda^k}{k!} \tag{3}$$

Remembering that TEs can involve adjacent pixels along both horizontal and vertical directions, the expected number of TEs of a perfect detector can be calculated as follows:

$$\begin{aligned} Expected\ No.\ of\ TE &= Probability\ of\ two\ non\text{-}zero\ pixels \cdot Total\ No.\ of\ 2\text{-}pix\ pairs \\ &= P(x \neq 0) \cdot P(x \neq 0) \cdot \big((H-1)W + H(W-1)\big) \\ &= \big(1 - P(x=0)\big)^2 \cdot (2HW - H - W) \\ &= \big(1 - e^{-\lambda}\big)^2 \cdot (2HW - H - W) \end{aligned} \quad (4)$$

where $x, H, W, \lambda$ are pixel value, height and width of the detector in unit of pixels, and the mean signal value, respectively. For a detector of size 512 ×1028 with a mean signal value of 0.027, the number of expected TE is ~ 746. The number of TEs from the real detector is 3891, which is much larger since it also contains false TEs. We also determined the number of TEs for the simulated perfect detector data and obtained 766 TEs. This is also apparent in the experimental and simulated images (**Fig 4c**): there are more large bright spots (corresponding to two or more adjacent bright pixels) in the experimental dataset, while the distribution of the spots is sparser.

While this effect acts as one of the main reasons for the modulation transfer function (MTF) degradation of direct electron detectors (Levin, 2021), it also undermines the performance of the denoiser. Even if the probability of a TE event is small, it still creates a strong signal correlation between adjacent pixels compared to the strength of the clean signal. Since the denoiser is designed to find correlations in the signal, it perceives the signal spread as a feature of the signal. To show this effect, the raw and denoised results from the experimental data of **fig. 4c** are presented in **fig 4d**. Patterns of cross-shape are formed after denoising, where the center pixel corresponds to one single bright pixel in the raw data. While the bright cross of the nearest neighbor pixels represents the signal spread across the interface, the center intensity represents the clean signal value inferred by the neural network.

It is possible to alleviate this problem. As shown in **fig 4a**, the lateral spread of the trajectory is less than the pixel size of the Dectris ELA detector (75 μm), i.e., the spreading is limited to neighboring pixels. Therefore, blinding the 8 neighboring pixels, in addition to the pixel of interest (i.e. 3 * 3 = 9 pixels in total), from the receptive field of the neural network (**fig 4e**) can avoid collecting this information, thus removing the cross-shape artifact. **Fig 4f-g** show a comparison between denoising results with and without blinding the neighboring pixels. For the blinding neighboring pixels result, the dark center cross-shape artifacts are suppressed, as shown in the EELS frame (**fig 4f**). Such artifacts are also present in the corresponding 1D spectra (**fig 4g**). While the denoised result without blinding neighboring pixels improves the SNR compared to the raw data, the spectrum has numerous spikes corresponding to the artifact, especially in flat background regions as the inset of **fig 4g** shows. All the experimental denoised results shown below were denoised with the 9-pixel blinding version of UDVD.

*3.3 Denoising Experimental Datasets – Core-Loss Spectrum and Near-Edge Fine Structures*

Here we present the denoised results of two core-loss EELS datasets: h-BN flake and gadolinium-doped ceria (GDC) nanoparticle. **Fig 5** shows the denoised result on a core-loss mapping of h-BN supported on amorphous carbon film. This set of results can showcase the capability of the denoiser to differentiate different components in the same dataset. The data is taken at 60 kV, and the total beam current was 25 pA. The acquisition time is 5ms per spectrum, and 1024 (32 × 32 pixels) spectra in total (~5 s total acquisition time). The energy dispersion was set to 0.1 eV/channel with 1028 channels, and the spectra were offset by 170 eV, thus covering an energy range from 170 eV to 272.8 eV. **Fig 5a** shows a comparison of the denoised spectra at three locations: on h-BN, on amorphous carbon, and in vacuum. As shown in the figure, signal from the h-BN shows the B K-edge with fine structures, while the spectrum from the amorphous carbon only shows a decaying background and the spectrum from vacuum is almost zero. **Fig 5b** shows a comparison of B K-edge between the denoised and raw data. There is a sign of the $\pi^*$ peak in the single spectrum of the raw data, however, the fine structures of the $\sigma^*$ peak are indistinguishable from the noise. The denoised spectrum resembles the average of 25 spectra, which show fine structures of the edge. It should be also noted that the denoised dataset has less noise than the 25-frame average spectrum. In other words, the denoiser can improve the SNR by at least 5 times in this case.

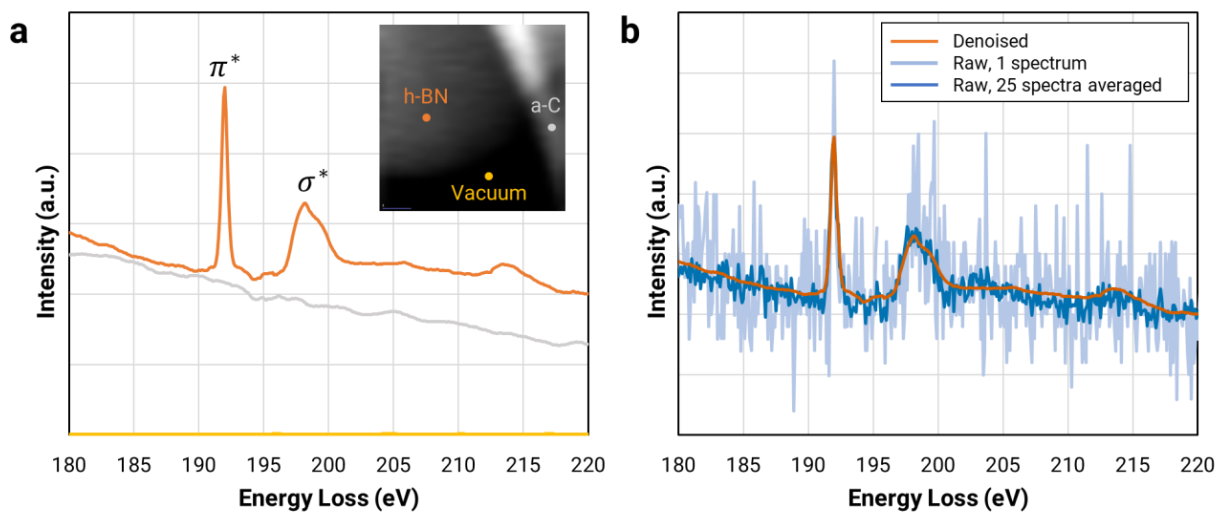

**Figure 5:** Denoising result of h-BN core-loss dataset. **a.** Comparison among denoised spectra from three different regions: h-BN (orange), amorphous carbon film (gray), and vacuum (yellow). π* and σ* peaks of boron are labelled accordingly. **b.** Comparison among single spectrum of raw (light blue) and denoised data (orange), and 25-frame averaged raw data (blue).

**Fig 6** presents the denoised result from an atomic resolution core-loss dataset from a GDC nanoparticle. These results demonstrate the capability of the denoiser to determine more subtle spectral changes. As **fig 6a** shows, the nanoparticle is tilted into (110) zone axis. The inset of **fig 6a** shows the scanning area from the SI. The data is taken at 100 kV, and the beam current was 60 pA. The acquisition time is 50ms per spectrum, and 1024 (32 × 32 pixels) spectra in total (~50 s total acquisition time). The energy dispersion was set to 0.5 eV/channel with 1028 channels, and the spectra were offset by 806 eV, thus covering an energy range from 806 eV to 1320 eV. **Fig 6b** compares the raw and denoised spectrum data. The denoised dataset has overall less noise compared to the raw spectra, and 25 spectra average. For the Ce $M_{4,5}$ edges, all the spectra show a lower $M_4$ peak compared to $M_5$ peak, which suggests a reduction of the Ce due to the beam irradiation (Garvie and Buseck, 1999). For signals much weaker than the Ce white lines, while the Gd $M_4$ edge is barely interpretable and the Ce $M_2$ edge is covered by noise even in the 25 spectra averaged, those two edges are clearly visible in the denoised spectrum.

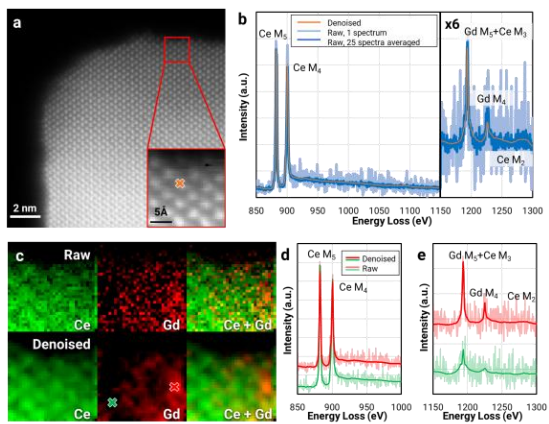

**Figure 6:** Denoising result of GDC core-loss dataset. **a.** HAADF image of the GDC. The figure inset presents the area EEL SI was performed. The orange cross marks the position of the spectrum in **fig 6b**. **b.** Comparison among single spectrum of raw (light blue) and denoised data (orange), and 64-frame averaged raw data (blue). The raw and denoised 1 spectrum data were taken at the orange cross in **a.** The 25 spectra averaged was taken from a box centered on the orange cross, which is the size of an atomic column. The data from energy range 1150 – 1300 eV was magnified to show the Gd peaks. **c.** SI of Ce and Gd. The red and green cross in Gd map shows the position where the spectra from **6d** and **e** were taken. **d.** Single raw (transparent color) and denoised (opaque color) spectra of Ce $M_{4,5}$ edges from strong (red) and weak (green) Gd signal region. **e.** Single raw and denoised spectra of Gd $M_{4,5}$ and Ce $M_{2,3}$ edges from Gd abundant and Gd scarce region.

**Fig 6c** shows the background subtracted intensity maps of Ce and Gd $M_{4,5}$ edges (see **supplement S4** for more details of data processing). Compared to the raw data, the denoised data has less noise. For the Gd map, the clusters of Gd atoms are more prominent compared to the raw data. To confirm our observation, the spectra taken from regions of weak and strong Gd signal (green and red cross in **fig 6c**, respectively) are compared in **fig 6d-e**. While the Ce $M_{4,5}$ signal is stronger in the green spectrum, the Gd $M_5$+Ce $M_3$ signal is weaker than the red spectrum. Also, the red spectrum has a much stronger Gd $M_4$ peak. Such fine structure information further confirms the integrity of the SI maps.

*3.4 Denoising Experimental Datasets – Vibrational Spectrum*

**Fig 7** demonstrates the denoising of an atomic resolution off-axis vibrational EELS dataset. This result showcases the capability of the denoiser to differentiate very small differences between spectra, even when the data has a large dynamic range. The dataset was acquired in an off-axis scheme reported by Hage et. al. (Hage et al., 2019), i.e., the center of the diffraction pattern was shifted 66 mrad away from the spectrometer entrance aperture center. The convergence semi-angle was set to 33 mrad, resulting in a probe size of 0.9 Å, and the collection semi-angle was set to 21 mrad. The total beam current was 25 pA after the insertion of monochromator slit, and $4.5 \times 10^{-3}$ pA passing through the spectrometer entrance aperture. The acquisition time was 100ms per spectrum, with a total of 2500 (50 × 50) spectra. The energy dispersion was set to 1 meV/channel. The full width at half maximum (FWHM) of the ZLP at the off-axis position was measured to be 17 meV.

A high-angle annular dark-field (HAADF) image of the sample is shown in **fig 7a**, where the red dashed box marks the mapping area. The inset of **fig 7a** shows the signal from HAADF detector when acquiring an energy-loss spectrum image in the off-axis geometry. Since the 000 beam was shifted off from the center of the EELS entrance aperture and part of it was captured by the HAADF detector, the signal is a bright field image. **Fig 7b** demonstrates a comparison between the raw and denoised single spectrum on and off the atomic column. While the spectra from two beam positions are indistinguishable in the raw data, the denoised spectra reveals a difference: the vibrational peaks taken on atomic column are stronger than the peaks in the off the atomic column, especially for low energy acoustic phonons. This result agrees with the result reported by Hage et al. (Hage et al., 2019). **Fig 7c** shows the comparison between raw and denoised energy-filtered maps. The ZLP maps of both the raw and denoised images are the same. For different phonon signals, detail in the maps from the raw data is mainly obscured by noise. After denoising, the maps show atomic resolution contrast. Also, the detailed contrast in the vibrational peak maps are all different, showing that the spatial variation in the scattering probabilities are different for these three peak regions.

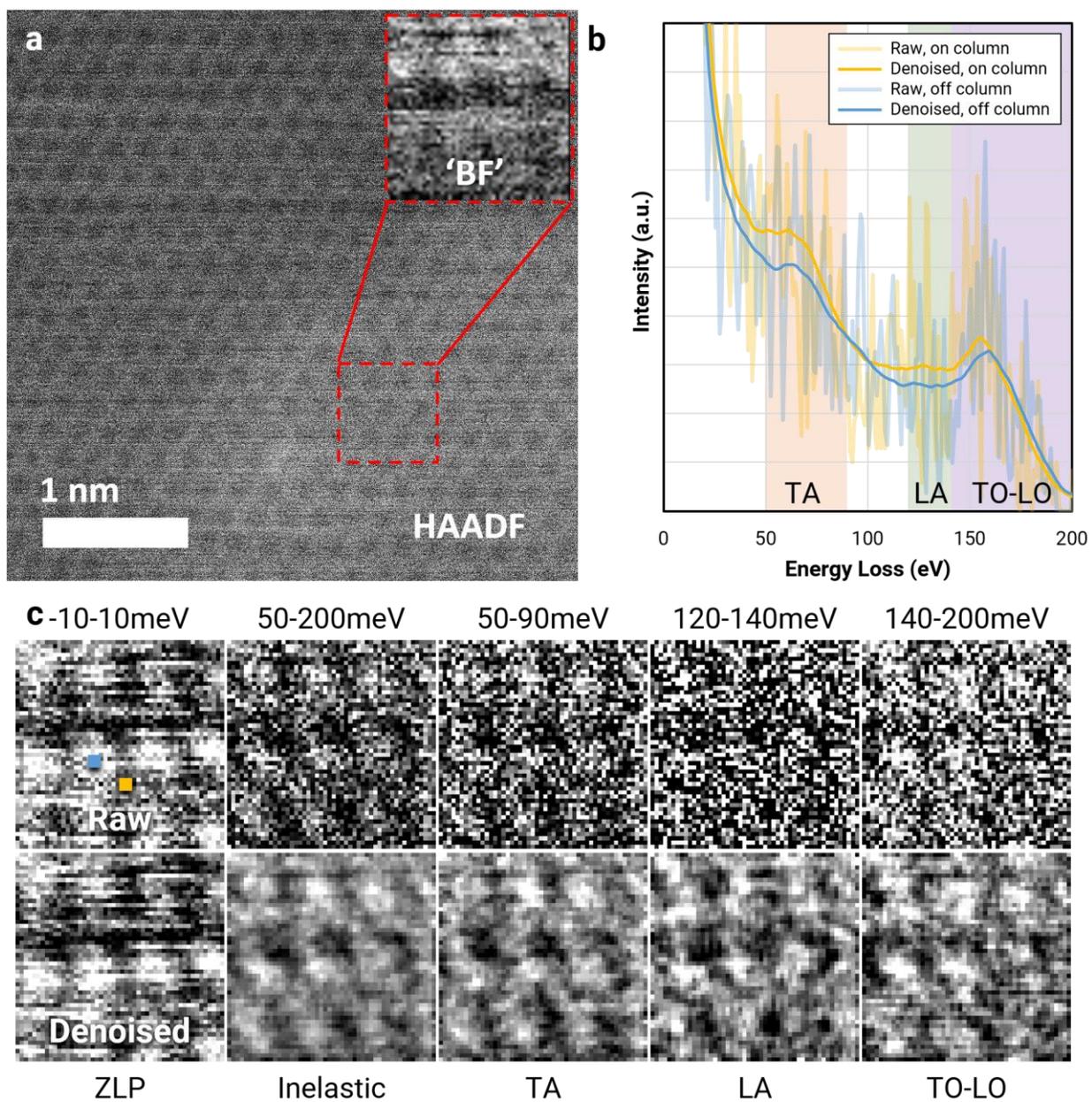

**Figure 7:** Denoising result of the vibrational loss dataset. **a.** Atomic resolution HAADF image of the sample. The red dashed box marks the place where the SI was taken. The figure inset is the signal collected from the HAADF detector when the spectroscopic data was taken, i.e. the beam was shifted off from the center of the EELS entrance aperture (off-axis), equivalent to a bright field image. **b.** Comparison between raw (transparent color) and denoised (opaque color) single spectrum on the atom column (yellow) and off the atom column (blue). The pixel of the spectrum is marked in **fig 8c** with corresponding color. Energy windows were marked with TA (50-90 meV), LA(120-140 meV), and TO-LO (140-200 meV) **c.** SIs from different energy loss features. The upper row is the raw data, and the bottom row is denoised data. The energy window and corresponding features are listed.

## 4. Conclusion

We have applied an unsupervised denoiser based on a deep neural network simulated and experimental EELS datasets. We modified the architecture of the network to achieve two objectives: making the network suitable for 4D dataset and avoiding generating artifacts from the detector. From the simulated dataset we showed an improvement of ~30 dB in PSNR. We also demonstrated various scenarios of the application of this method to experimental datasets, including core-loss datasets and a vibrational loss dataset. Through the core-loss dataset, we showed that the neural network can easily differentiate various composites in one dataset, also determine fine structure details in a spectrum. For the vibrational loss data, we showed that the neural network can distinguish small differences between spectrum, thus achieving atomic resolution. We believe this powerful new tool can open horizons for new applications of EELS and developments of new EELS techniques.

## CRediT authorship contribution statement

**Yifan Wang:** Investigation, Formal analysis, Software, Methodology, Validation, Visualization, Writing – Original Draft, Writing – Review& Editing. **Mai Tan**: Investigation, Writing – Review& Editing. **Carlos Fernandez-Granda:** Methodology, Writing – Review& Editing, Funding acquisition. **Peter A. Crozier:** Writing – Review& Editing, Funding acquisition.

## Acknowledgements

We gratefully acknowledge support of NSF grants to ASU (CHE-2109202, OAC-1940263, 2104105) and NYU (HDR-1940097 and OAC-2103936). We also gratefully acknowledge the use of ASU's John M. Cowley Center for High Resolution Electron Microscopy, and the use of Sol supercomputer at ASU (Jennewein et al., 2023). We also gratefully thank the insightful discussions with Dr. Ben Plotkin-Swing at Bruker Co. and Dr. Barnaby Levin at Direct Electron L.P. on the artifacts from the detector. We would like to acknowledge Prof. Seth Ariel Tongay and Patrick Hays at ASU for providing the h-BN sample.

**Supplementary Materials**

*S1 Adaptation of UDVD to a 4D dataset*

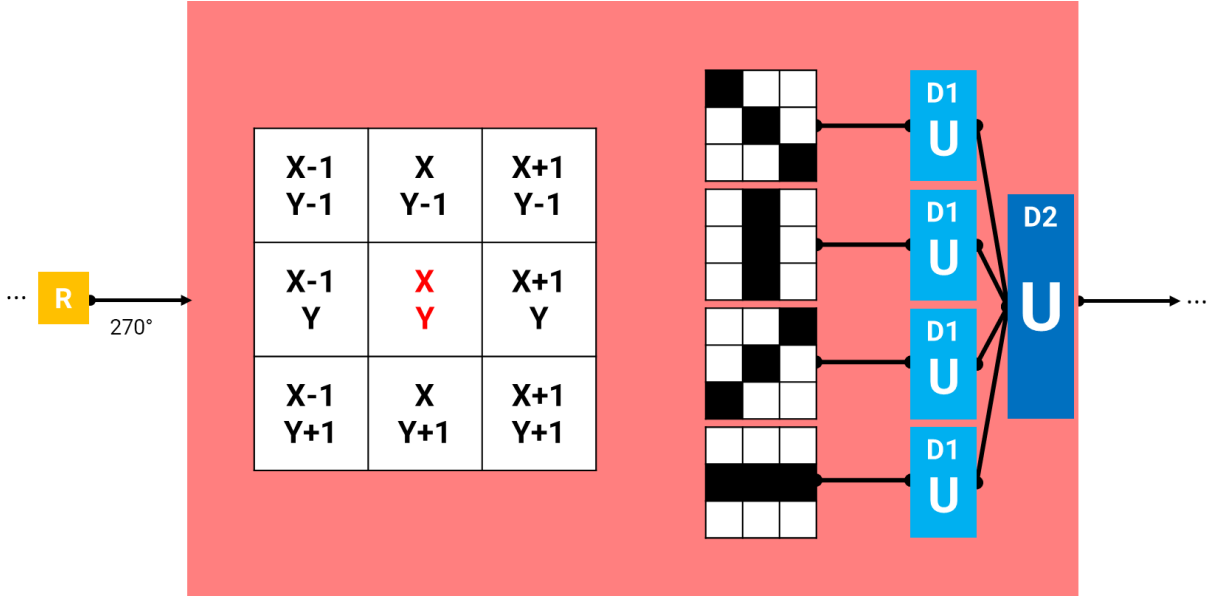

**Figure S1:** Detailed modifications to UDVD for a 4D dataset. The four rotated branches are changed, while the rest of the architecture remains the same. For clarification purposes, only one of the branches is shown here (270° rotation). The 9 squares in the box on the left represent the 9 frames used to estimate the values in frame (X, Y). The black squares in the boxes represent the input to each D1 U-net, i.e., from top to bottom: neighbors in one diagonal, vertical neighbors, neighbors in the other diagonal, and horizontal neighbors.

While the original UDVD only uses five frames as an input, we used 9 frames for the 4D dataset, i.e., to estimate pixel values in frame (X, Y), frames (X-1, Y-1) to (X+1, Y+1) are used as inputs. In this case, the middle pixel has four sets of neighbors: horizontal, vertical, and two diagonals (**Figure S1**). To incorporate this feature, we added an additional U-net component. Therefore, there are four U-net components in layer D1.

*S2 Relationship between SNR and PSNR*

We assume the signal follows Poisson statistics, i.e., for a signal of mean $\sigma$, the mean squared error is $\sigma$. For an $n$-fold signal increase, the corresponding increase of PSNR is:

$$PSNR - PSNR_0 = 10 \cdot \log_{10}\left(\frac{(n \cdot MAX_I)^2}{n \cdot MSE}\right) - 10 \cdot \log_{10}\left(\frac{MAX_I^2}{MSE_I}\right) = 10 \cdot \log_{10} n$$

i.e., an increase of $x$ dB in PSNR corresponds to a signal increase of $10^{0.1x}$ time.

*S3 Core-loss Atomic Resolution EELS Map Simulation*

A multilayer h-BN atomic model was built with point defects, e.g., adatoms or vacancies (**fig S1a**). The STEM probe profile is simplified to a 2D gaussian with a FWHM of 1.5 Å. A square scan with width of 32 pixels, corresponding to an area of 8 × 8 Å² was created. To determine the strength of the elemental signal at each probe position $(x, y)$, two intensity masks corresponding to B and N were calculated by convolving the probe profile centered at $(x, y)$ with number of atoms at various positions (**fig S1b**).

To obtain the absolute loss probability, a spectrum was acquired from h-BN with both core-loss and low-loss for calculation of sample thickness. The spectrum was divided into 3 components through the power law background subtraction: B K-edge, N K-edge, and background. The loss probability of one B and N atom as functions of energy loss were estimated through normalizing the elemental edge with total intensity and divided by number of atoms, derived from the sample thickness. The remaining background was normalized to the thickness of the atomic model. By combining the intensity masks and absolute loss probability information, a 3D SI $I(x, y, \Delta E)$ was constructed (**fig S1c**).

To match the 2D signal on the spectrometer detector $(q, \Delta E)$, the spectrum signal at each position were convolved with the point spread function:

$$I(x, y, \Delta E) * PSF(q, \Delta E) = g(x, y, \Delta E, q),$$

$$PSF(q, \Delta E) = \begin{cases} 2\sqrt{q_{max}^2 - q^2}, & |q| \leq q_{max}, \Delta E = 0, \\ 0, & elsewhere \end{cases}$$

where $q, q_{max}$ are the non-dispersive (momentum) axis coordinate and half-width of the probe in non-dispersive direction on the detector, respectively. The PSF above corresponds to the optics of a perfect spectrometer, i.e., focused on dispersive direction, and no spectrometer aberration. The result of the convolution process, $g(x, y, \Delta E, q)$, is taken as the ground truth. Various combinations of probe current and exposure time were set and converted to number of electrons arriving. Noisy 4D SIs $G(x, y, \Delta E, q)$ were generated through a Poisson process.

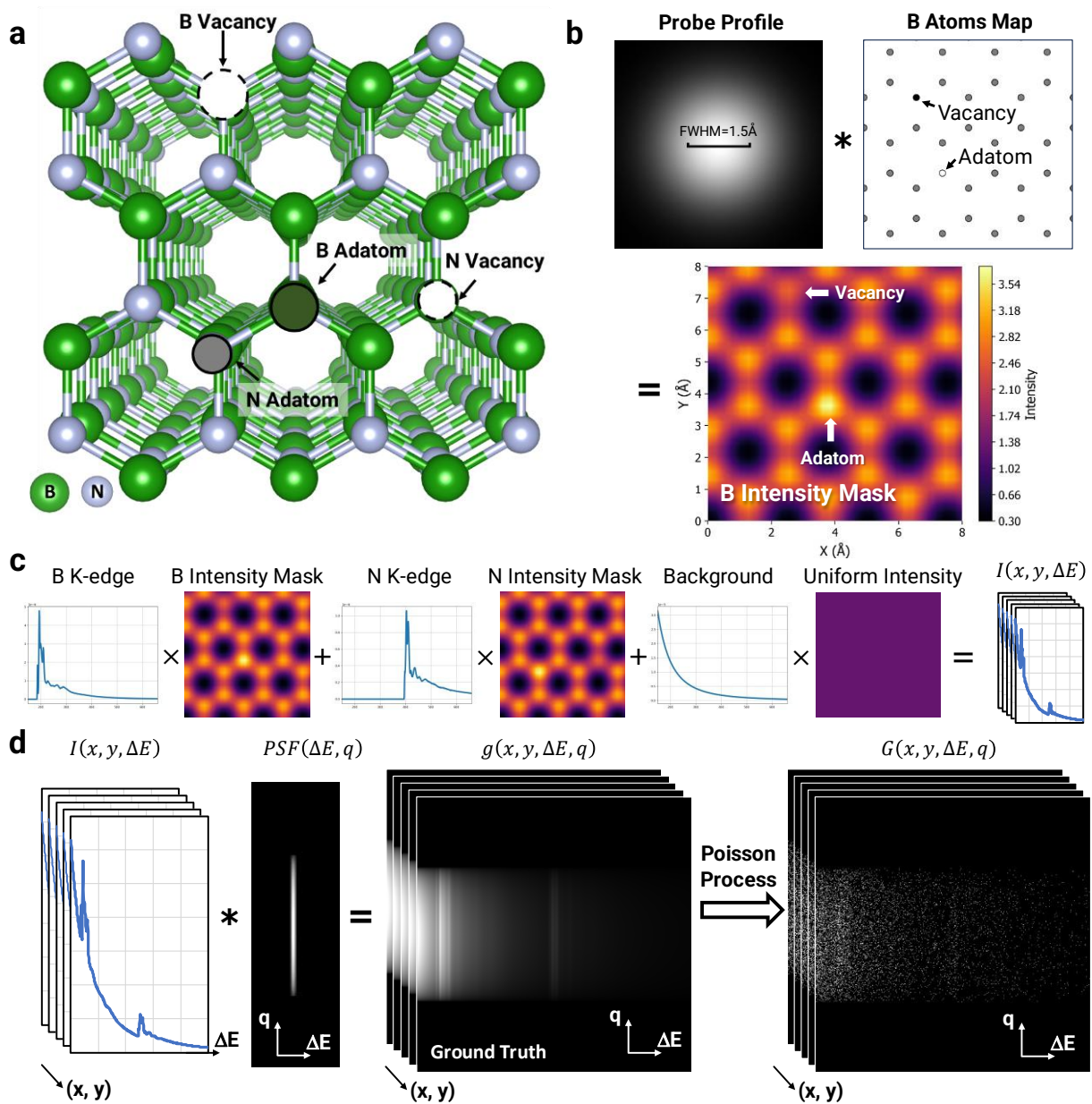

**Figure S2 a.** The atomic model of a 10-layer h-BN viewed from [0001] direction. The green atoms represent B and gray atoms represent N. Defects of adatoms and vacancies for both B and N are created and marked by black arrows. **b.** Schematic diagram of B elemental signal strength map. The map was constructed by the convolution of the probe intensity profile with the B atom map. The position of the extra atom and vacancy (white arrows) agrees with the atomic model. **c.** Constructing the 3D SI. The dataset was a summation of the product of EELS signal and their corresponding intensity mask. **d.** Constructing the 4D dataset. The dataset $g(x,y,\Delta E, q)$ was generated through the convolution of the spectra $I(x,y,\Delta E)$ and the spectrometer point spread function $PSF(\Delta E, q)$.

*S4 Ce Gd signal processing*

Here we show how we process the Ce and Gd core-loss signal in section 3.3. Major edges for Ce are $M_4$ (901 eV), $M_5$ (883 eV) and minor edges are $M_2$ (1273 eV), $M_3$ (1185 eV). For Gd, major edges are $M_4$ (1217 eV), $M_5$ (1185 eV) and minor edges are $M_2$ (1688 eV), $M_3$ (1544 eV). We applied power law background subtraction to extract signal of Ce $M_{45}$ edges and Gd $M_{45}$ edges. The windows we used to fit the background are 800 to 840 eV for Ce $M_{45}$, and 1065 to 1165 eV for Gd $M_{45}$. The signal window was kept at 120 eV for both elemental edges, which is 840 to 960 eV for Ce, and 1165 to 1285 eV for Gd.

Notice that Gd $M_5$ edge overlaps with Ce $M_3$ edge, and Ce $M_2$ edge is also included in the Gd signal window. Thus, to extract Gd signal, we need to subtract the Ce $M_{23}$ edges. Although the ratio of Ce M4 over Ce M5 does vary as an indication of different levels of reduction of the sample, we have observed the ratio of Ce M45 signal and Ce M23 signal for those different areas remained a constant of $10.3 \pm 0.1$ regardless of different thickness and oxidation state (based on 20 spectrum measurement from different regions of the sample).

Knowing the ratio between the Ce M45 and Ce M23 signals can help us determine the contribution of Ce M23 in the overlap region. The ratio between Ce $M_{45}$ ($I_{Ce\,M_{45}}$) and Ce $M_{23}$ ($I_{Ce\,M_{23}}$) signal written as:

$$\frac{I_{Ce\,M_{45}}}{I_{Ce\,M_{23}}} = Constant = 10.3 \pm 0.1$$

Using the constant ratio relationship, we can calculate Gd $M_{45}$ signal, $I_{Gd\,M_{45}}$, as:

$$I_{Gd\,M_{45}} = I_{overlap} - I_{Ce\,M_{23}}$$
$$= I_{overlap} - I_{Ce\,M_{45}}/10.3$$

where $I_{overlap}$ is the signal from 1165 to 1285 eV after background subtraction.